\documentclass{article}
\usepackage{amssymb}
\usepackage{amsmath}

\input{tcilatex}

\begin{document}

\title{Revisiting the Neyman-Scott model: \\
an Inconsistent MLE or an Ill-defined Model?}
\author{\textbf{Aris Spanos} \\
%EndAName
Department of Economics,\\
Virginia Tech, USA}
\date{January 2013}
\maketitle

\begin{abstract}
The Neyman and Scott (1948) model is widely used to demonstrate a serious
weakness of the Maximum Likelihood (ML) method: it can give rise to
inconsistent estimators. The primary objective of this paper is to revisit
this example with a view to demonstrate that the culprit for the
inconsistent estimation is not the ML method but an ill-defined statistical
model. It is also shown that a simple recasting of this model renders it
well-defined and the ML method gives rise to consistent and asymptotically
efficient estimators.
\end{abstract}

\newpage 

\section{Introduction\protect\vspace*{-0.1in}}

Despite claims for priority by a number of different authors, Maximum
Likelihood (ML) was first articulated as a general method of estimation in
the context of a parametric statistical model by Fisher (1922). It took
several decades to establish the regularity conditions needed to ensure the
key asymptotic properties of ML Estimators (MLE), such as efficiency and
consistency (Cramer, 1946, Wald, 1949), but since then ML has dominated
estimation in frequentist statistics; see Stigler (2007), Hald (2007) for
this history.

Several counterexamples were proposed in the 1940s and 1950s raising doubts
about the generality of the ML method. These counterexamples include
Hodges's local superefficient estimator (Le Cam, 1953), the mixture of two
Normal distributions (Cox, 2006)) and the inconsistent MLE example proposed
by the Neyman and Scott (1948) model. Commenting on these examples Stigler
(2007), p. 613, argued that none are considered serious enough to undermine
the credibility of the ML method, and singled out the last example:

\textsf{\textquotedblleft The Wald-Neyman-Scott example was of more
practical import, and still serves as a warning of what might occur in
modern highly parameterized problems, where the information in the data may
be spread too thinly to achieve asymptotic consistency.\textquotedblright\ }

The primary objective of this note is to revisit the Neyman-Scott example
with a view to unpack Stigler's assessment by demonstrating that the real
culprit for the inconsistent estimator is not the ML method, as such, but an
ill-defined statistical model. It is also shown that a simple recasting of
this model renders it well-defined and the ML method gives rise to
consistent and asymptotically efficient estimators.

\section{The Neyman-Scott model}

The quintessential example used to demonstrate that ML might give rise to
inconsistent estimators is \textit{the Neyman-Scott model}\textbf{:}%
\[
\fbox{%
\begin{tabular}{l}
$\left. 
\begin{array}{c}
X_{it}=\mu _{t}+\varepsilon _{it},\vspace*{0.08in} \\ 
\varepsilon _{it}\backsim \text{\textsf{NIID}}(0,\sigma ^{2})=0,%
\end{array}%
\right\} \ i\mathbf{=}1,2,\ t\mathbf{=}1,2,...,n,...$%
\end{tabular}%
}
\]%
which can be viewed as a \textsf{simple time effects panel data model}.

The underlying distribution of the observable random variables $\left(
X_{1t},X_{2t}\right) $ is bivariate Normal Independent, but \textit{not}
Identically Distributed: 
\[
\mathbf{X}_{t}\text{:=}\left( 
\begin{array}{c}
X_{1t} \\ 
X_{2t}%
\end{array}%
\right) \backsim \text{\textsf{NI}}\left( \left( 
\begin{array}{c}
\mu _{t} \\ 
\mu _{t}%
\end{array}%
\right) ,\left( 
\begin{array}{cc}
\sigma ^{2} & 0 \\ 
0 & \sigma ^{2}%
\end{array}%
\right) \right) ,\ t\mathbf{=}1,2,...,n,...
\]

In light of the fact that the non-ID\ assumption implies that this model
suffers from the \textit{incidental parameter problem}, in the sense that
the unknown parameters $(\mu _{1},\mu _{2},...,\mu _{n})$ increase with the
sample size, the latter are viewed as \textit{nuisance} parameters, and\ $%
\sigma ^{2}$ as the only parameter of interest.

\section{Maximum Likelihood Estimation?}

Despite the incidental parameter problem the Neyman-Scott model continues to
be used as a counterexample to the ML method. That is, the literature
ignores the incidental parameter problem, and defines the distribution of
the sample to be:\vspace*{-0.08in}%
\[
\begin{array}{cl}
f(\mathbf{x};\mathbf{\theta }) & \mathbf{=}\prod\limits_{t\mathbf{=}%
1}^{n}\prod\limits_{i\mathbf{=}1}^{2}\frac{1}{\sigma \sqrt{2\pi }}e^{\left\{
-\frac{1}{2\sigma ^{2}}(x_{it}-\mu _{t})^{2}\right\} }\mathbf{=}%
\prod\limits_{t\mathbf{=}1}^{n}\frac{1}{2\pi \sigma ^{2}}e^{\left\{ -\frac{1%
}{2\sigma ^{2}}[(x_{1t}-\mu _{t})^{2}+(x_{2t}-\mu _{t})^{2}]\right\} }%
\end{array}%
\vspace*{-0.08in}
\]%
This gives rise to the `log-likelihood function':%
\[
\begin{array}{cl}
\ln L(\mathbf{\theta };\mathbf{x}) & \mathbf{=}-n\ln \sigma ^{2}\mathbf{-}%
\frac{1}{2\sigma ^{2}}\sum_{t\mathbf{=}1}^{n}[(x_{1t}\mathbf{-}\mu
_{t})^{2}+(x_{2t}\mathbf{-}\mu _{t})^{2}].%
\end{array}%
\]%
The `Maximum Likelihood Estimators' (MLE)\ are supposed to be derived by
solving the first-order conditions:\vspace*{-0.08in}%
\[
\begin{array}{ll}
\frac{\partial \ln L(\mathbf{\theta };\mathbf{x})}{\partial \mu _{t}} & 
\mathbf{=}\frac{1}{\sigma ^{2}}[(x_{1t}-\mu _{t})+(x_{2t}-\mu _{t})]\mathbf{=%
}0,%
\end{array}%
\vspace*{-0.08in}
\]%
\[
\begin{array}{ll}
\frac{\partial \ln L(\mathbf{\theta };\mathbf{x})}{\partial \sigma ^{2}} & 
\mathbf{=-}\frac{n}{\sigma ^{2}}\mathbf{+}\frac{1}{2\sigma ^{4}}\sum_{t%
\mathbf{=}1}^{n}[(x_{1t}-\mu _{t})^{2}+(x_{2t}-\mu _{t})^{2}]\mathbf{=}0,%
\end{array}%
\vspace*{-0.08in}
\]%
giving rise to:\vspace*{-0.09in} 
\[
\begin{array}{l}
\begin{array}{c}
\widehat{\mu }_{t}\mathbf{=}\frac{1}{2}(X_{1t}+X_{2t}),\ t\mathbf{=}%
1,2,...,n,%
\end{array}%
\medskip \\ 
\begin{array}{c}
\widehat{\sigma }^{2}\mathbf{=}\frac{1}{2n}\sum\limits_{t\mathbf{=}%
1}^{n}[(X_{1t}-\widehat{\mu }_{t})^{2}+(X_{2t}-\widehat{\mu }_{t})^{2}]%
\mathbf{=}\frac{1}{n}\sum\limits_{t\mathbf{=}1}^{n}s_{t}^{2},%
\end{array}%
\end{array}%
\]%
where $s_{t}^{2}$\textbf{$=$}$\frac{1}{2}[(X_{1t}-\widehat{\mu }%
_{t})^{2}+(X_{2t}-\widehat{\mu }_{t})^{2}].$

Notice that for $\ln L(\mathbf{\theta };\mathbf{x)}$, $\widehat{\mathbf{%
\theta }}_{MLE}\mathbf{:}$\textbf{$=$}$(\widehat{\mu }_{t},\hat{\sigma}^{2},t%
\mathbf{=}1,2,...,n)$ is a maximum since the second derivatives at $\mathbf{%
\theta =}\widehat{\mathbf{\theta }}$ are:\vspace*{-0.08in}%
\[
\begin{array}{ll}
\left. \frac{\partial ^{2}\ln L(\mathbf{\theta };\mathbf{x})}{\partial \mu
_{t}^{2}}\right\vert _{\mathbf{\theta =\widehat{\theta }}_{MLE}}\mathbf{=} & 
\left. -\left( \frac{2}{\sigma ^{2}}\right) \right\vert _{\mathbf{\theta =%
\widehat{\theta }}_{MLE}}\mathbf{=}-\frac{2}{\hat{\sigma}^{2}}<0,%
\end{array}%
\vspace*{-0.08in}
\]

\[
\begin{array}{ll}
\left. \frac{\partial ^{2}\ln L(\mathbf{\theta };\mathbf{x})}{\partial
\sigma ^{2}\partial \mu _{t}}\right\vert _{\mathbf{\theta =\widehat{\theta }}%
_{MLE}}\mathbf{=} & \left. -\frac{1}{\sigma ^{4}}\sum_{t\mathbf{=}%
1}^{n}[(x_{1t}-\mu _{t})+(x_{2t}-\mu _{t})]\right\vert _{\mathbf{\theta =%
\widehat{\theta }}_{MLE}}\mathbf{=}0%
\end{array}%
\vspace*{-0.08in}
\]

\[
\begin{array}{ll}
\left. \frac{\partial ^{2}\ln L(\mathbf{\theta };\mathbf{x})}{\partial
\sigma ^{4}}\right\vert _{\mathbf{\theta =\widehat{\theta }}_{MLE}}\mathbf{=}%
\left. \frac{n}{\sigma ^{4}}\mathbf{-}\frac{1}{\sigma ^{6}}\sum_{t\mathbf{=}%
1}^{n}[(x_{1t}-\mu _{t})^{2}+(x_{2t}-\mu _{t})^{2}]\right\vert _{\mathbf{%
\theta =\widehat{\theta }}_{MLE}}\mathbf{=}-\frac{n}{\hat{\sigma}^{4}}<0, & 
\end{array}%
\vspace*{-0.08in}
\]

and thus:\vspace*{-0.08in}%
\[
\begin{array}{c}
\left. \left( \frac{\partial ^{2}\ln L(\mathbf{\theta };\mathbf{x})}{%
\partial \mu _{t}^{2}}\right) \left( \frac{\partial ^{2}\ln L(\mathbf{\theta 
};\mathbf{x})}{\partial \sigma ^{4}}\right) -\left( \frac{\partial ^{2}\ln L(%
\mathbf{\theta };\mathbf{x})}{\partial \sigma ^{2}\partial \mu _{t}}\right)
\right\vert _{\mathbf{\theta =\widehat{\theta }}_{MLE}}>0.%
\end{array}%
\vspace*{-0.08in}
\]

The commonly used argument against the ML method is that since:%
\[
\begin{array}{r}
E(\widehat{\mu }_{t})\mathbf{=}\mu _{t}\text{ and }E(s_{t}^{2})\mathbf{=}%
\frac{1}{2}\sigma ^{2},%
\end{array}%
\]%
it follows that the `MLE'\ $\widehat{\sigma }^{2}$ is both\textbf{\ }\textit{%
biased} and \textit{inconsistent} because: 
\[
\begin{array}{rrr}
E(\widehat{\sigma }^{2})\mathbf{=}\frac{1}{n}\sum_{t\mathbf{=}%
1}^{n}E(s_{t}^{2})\mathbf{=}\frac{1}{2}\sigma ^{2},\  & \text{and} & 
\widehat{\sigma }^{2}\overset{a.s.}{\rightarrow }\frac{1}{2}\sigma ^{2},%
\end{array}%
\]%
since the bias $E(\widehat{\sigma }^{2})\mathbf{-}\sigma ^{2}$\textbf{$=$}$%
\mathbf{-}\frac{1}{2}\sigma ^{2}$ does \textit{not} go to $0$ as $n\mathbf{%
\rightarrow }\infty .$

A moment's reflection reveals that the inconsistency argument is ill-thought
out. The is because the incidental parameter problem renders $\widehat{\mu }%
_{t}$\textbf{$=$}$\frac{1}{2}(X_{1t}+X_{2t})$\textit{\ an inconsistent}
estimator of $\mu _{t},$ for $t$\textbf{$=$}$1,2,...,n;$ there are only two
observations for each $\widehat{\mu }_{t},$ and thus their variance $Var(%
\widehat{\mu }_{t})$\textbf{$=$}$\frac{1}{2}\sigma ^{2}$ does not go to zero
as $n\rightarrow \infty .$

What the critics of the ML method do not appreciate enough is the fact that
treating the unknown parameters $(\mu _{1},\mu _{2},...,\mu _{n})$ as
incidental and designating $\sigma ^{2}$ the only parameter of interest,
does not let the statistician `off the hook'. This is because the parameter
of interest $\sigma ^{2}$\textbf{$=$}$E(X_{it}\mathbf{-}\mu _{t})^{2},$
defining the variation around $\mu _{t}$, invokes the incidental parameters.
Put more intuitively, when the data come in the form of $\mathbf{Z}_{0}%
\mathbf{:=}(\mathbf{z}_{1},\mathbf{z}_{2},...,\mathbf{z}_{n}),$ $\mathbf{z}%
_{t}\mathbf{:=}(x_{1t},x_{2t})$, one can get to $\sigma ^{2}$ via $\mu _{t}$%
, and using $\widehat{\mu }_{t}$ leads to problems because it is an
inconsistent estimator. In that sense (Severini, 2000):

\textsf{\textquotedblleft ... this model falls outside of the general
framework we are considering since the dimension of the parameter }$(\mu
_{1},\mu _{2},...,\mu _{n},\sigma ^{2})$\textsf{\ depends on the sample
size.\textquotedblright\ (p. 108)}\newline
That is, calling $\widehat{\sigma }^{2}\mathbf{=}\frac{1}{2n}\sum\limits_{t%
\mathbf{=}1}^{n}[(X_{1t}\mathbf{-}\widehat{\mu }_{t})^{2}\mathbf{+}(X_{2t}%
\mathbf{-}\widehat{\mu }_{t})^{2}]$ a MLE is highly misleading since the ML
method was never meant to be applied to statistical models whose number of
unknown parameters increases with the sample size $n.$

In truth, one should be very skeptical of any method of estimation which
yields consistent estimators in cases where the statistical model in
question is \textit{ill-defined}\textbf{,} as in the case of the incidental
parameter problem. Hence, the more interesting question should be:\vspace*{%
-0.15in}%
\[
\begin{tabular}{l}
\textsf{why would the ML method yield a consistent estimator of} $\sigma
^{2}?$%
\end{tabular}%
\vspace*{-0.08in}
\]%
The fact that the ML method does \textit{not} yield a consistent estimator
of $\sigma ^{2}$ should count in its favor not against it! To paraphrase
Stigler's quotation: the ML method `\textsf{warns the modeler that the
information in the data has been spread too thinly'.}

\section{Recasting the original Neyman-Scott model}

The question that naturally arises is: can one respecify the above
statistical model to render it well-defined but retaining the parameter of
interest? The answer is surprisingly straightforward. Since the incidental
parameter problem arises because of the unknown but \textit{t-varying} means 
$(\mu _{1},\mu _{2},...,\mu _{n})$, one can re-specify the original
bivariate model into a univariate\textbf{\ }\textit{simple Normal} (one
parameter), using the transformation:\vspace*{-0.08in}%
\[
\fbox{$%
\begin{array}{c}
Y_{t}\mathbf{=}\frac{1}{\sqrt{2}}(X_{1t}-X_{2t})\backsim \text{\textsf{NIID}}%
\left( 0,\sigma ^{2}\right) ,\ t\mathbf{=}1,2,...,n,...,%
\end{array}%
$}
\]%
\[
\begin{array}{c}
E(Y_{t})\mathbf{=}\frac{1}{\sqrt{2}}E(X_{1t}-X_{2t})\mathbf{=}\frac{1}{\sqrt{%
2}}(\mu _{t}-\mu _{t})\mathbf{=}0,\medskip \\ 
Var(Y_{t})=\frac{1}{2}[Var(X_{1t})+Var(X_{2t})]\mathbf{=}\sigma ^{2}.%
\end{array}%
\]%
This is a sensible thing to do because taking the difference eliminates the 
\textit{nuisance} parameters $(\mu _{1},\mu _{2},...,\mu _{n})$, without
affecting the parameter of interest. For this simple Normal model, the MLE
for $\sigma ^{2}$ is: $%
\begin{array}{c}
\widehat{\sigma }_{MLE}^{2}\mathbf{=}\frac{1}{n}\sum\nolimits_{t\mathbf{=}%
1}^{n}Y_{t}^{2},%
\end{array}%
\medskip \newline
$which it is\textsf{\ }\textit{unbiased, fully efficient} and \textit{%
strongly consistent:}\vspace*{-0.08in}%
\[
\begin{array}{c}
E(\widehat{\sigma }_{MLE}^{2})\mathbf{=}\sigma ^{2},\ Var(\widehat{\sigma }%
_{MLE}^{2})\mathbf{=}\frac{2\sigma ^{4}}{n},\ \widehat{\sigma }_{MLE}^{2}%
\overset{a.s.}{\rightarrow }\sigma ^{2}.%
\end{array}%
\vspace*{-0.08in}
\]

\underline{Notes}:

(i) The above recasting of the Neyman-Scott model can be easily extended to
the case $\left( X_{1t},X_{2t},...,X_{mt}\right) ,\ 2\leq m<n.$

(ii) Hald (2007), p. 182-3 offers an alternative, highly original, way to
sidestep the incidental parameter problem using Fisher's two stage ML method.

\section{Conclusion}

The main conclusion from the above discussion is that when the ML method
gives rise to inconsistent estimators, the modeler should take a closer look
at the assumed statistical model; chances are, it is ill-defined. This is
particularly true in the case where the assumed model suffers from the
incidental parameter problem. In such cases the way forward is to recast the
original model to render it well-defined and then apply the ML method. This
argument is illustrated above using the\ Neyman-Scott (1948) model. \vspace*{%
-0.1in}


\begin{thebibliography}{9}
\bibitem{} Cox, D. R. (2006), \textit{Principles of Statistical Inference},
Cambridge University Press, Cambridge.\vspace*{-0.1in}

\bibitem{} Cramer, H. (1946), \textit{Mathematical Methods of Statistics},
Princeton University Press, Princeton, NJ.\vspace*{-0.1in}

\bibitem{} Fisher, R. A. (1922), \textquotedblleft On the mathematical
foundations of theoretical statistics\textquotedblright , \textit{%
Philosophical Transactions of the Royal Society A}, \textbf{222}: 309-368.%
\vspace*{-0.1in}

\bibitem{} Hald, A. (2007), \textit{A History of Parametric Statistical
Inference from Bernoulli to Fisher, 1713-1935}, Springer, NY.\vspace*{-0.1in}

\bibitem{} Le Cam, L. (1953), \textquotedblleft On some asymptotic
properties of maximum likelihood estimates and related Bayesian
estimates,\textquotedblright\ \textit{University of California Publications
in Statistics}, 1: 277-330.\vspace*{-0.1in}

\bibitem{} Neyman, J. and E. L. Scott (1948), \textquotedblleft Consistent
estimates based on partially consistent observations,\textquotedblright\ 
\textit{Econometrica}, 16: 1-32.\vspace*{-0.1in}

\bibitem{} Severini, T. A. (2000), \textit{Likelihood Methods in Statistics}%
, Oxford University Press, Oxford.\vspace*{-0.1in}

\bibitem{} Stigler, S. M. (2007), \textquotedblleft The Epic Story of
Maximum Likelihood,\textquotedblright\ \textit{Statistical Science}, 22:
598-620.\vspace*{-0.1in}

\bibitem{} Wald, A. (1949), \textquotedblleft Note on the consistency of the
maximum likelihood estimate,\textquotedblright\ \textit{Annals of
Mathematical Statistics}, 20: 595-601.
\end{thebibliography}
\end{document}